\documentclass[12pt,a4paper]{article}
\usepackage{amsmath}
\usepackage{amsfonts}
\usepackage{cite}

\newcommand{\Eqn}[1]{&\hspace{-0.5em}#1\hspace{-0.5em}&}
\renewcommand{\[}{\begin{equation}}
\renewcommand{\]}{\end{equation}}

\newcommand{\bbC}{{\mathbb C}}
\newcommand{\bbR}{{\mathbb R}}
\newcommand{\bbZ}{{\mathbb Z}}

\newcommand{\grp}[1]{\mathrm{#1}}

\newcommand{\nvec}[1]{\boldsymbol{#1}}

\newcommand{\bw}{{\bar{w}}}
\newcommand{\bz}{{\bar{z}}}
\newcommand{\balpha}{\bar{\alpha}}
\newcommand{\bdelta}{{\bar{\delta}}}
\newcommand{\bnu}{{\bar{\nu}}}
\newcommand{\bA}{{\bar{A}}}

\newcommand{\iDelta}{\mathnormal{\Delta}}
\newcommand{\iLambda}{\mathnormal{\Lambda}}
\newcommand{\iPhi}{\mathnormal{\Phi}}
\newcommand{\iPsi}{\mathnormal{\Psi}}

\newcommand{\iOmega}{\mathnormal{\Omega}}

\newcommand{\vm}{\nvec{m}}
\newcommand{\vu}{\nvec{u}}
\newcommand{\vz}{\nvec{z}}
\newcommand{\vX}{\nvec{X}}
\newcommand{\vU}{\nvec{U}}
\newcommand{\vV}{\nvec{V}}
\newcommand{\vD}{\nvec{D}}
\newcommand{\vDelta}{\nvec{\iDelta}}
\newcommand{\vzero}{\nvec{0}}

\newcommand{\cL}{\mathcal{L}}
\newcommand{\cU}{\mathcal{U}}
\newcommand{\cV}{\mathcal{V}}

\newcommand{\hvarphi}{{\hat{\varphi}}}
\newcommand{\hC}{{\hat{C}}}
\newcommand{\hU}{{\hat\cU}}
\newcommand{\hV}{{\hat\cV}}

\newcommand{\tU}{{\tilde\cU}}
\newcommand{\tV}{{\tilde\cV}}
\newcommand{\tPhi}{{\tilde\iPhi}}
\newcommand{\tmu}{{\tilde\mu}}

\newcommand{\unitmatrix}{{\boldsymbol{1}}}

\newcommand{\varth}{\vartheta}

\renewcommand{\thesection}
  {\arabic{section}.\hspace{-.5em}}

\makeatletter
\renewcommand\section{
  \@startsection{section}{3}{\z@}%
  {-3.25ex\@plus -1ex \@minus -.2ex}%
  {1.5ex \@plus .2ex}%
  {\normalfont\normalsize\bfseries\mathversion{bold}}}
\makeatother

\makeatletter \@addtoreset{equation}{section} \makeatother
\renewcommand{\theequation}{\arabic{section}.\arabic{equation}}

\makeatletter
\renewcommand{\appendix}{
  \renewcommand{\thesection}{Appendix \Alph{section}.\hspace{-.5em}}
\@addtoreset{equation}{subsection}
\renewcommand{\theequation}{\Alph{section}.\arabic{equation}}
\setcounter{section}{0}}
\makeatother

\def\eqb         {  \begin{eqnarray}  }
\def\eqe           {  \end{eqnarray}  }
\def\nn               {  \nonumber  }
\newcommand{\vn}{\nvec{n}}

\begin{document}
%
\def\papertitlepage{\baselineskip 3.5ex \thispagestyle{empty}}
\def\preprinumber#1#2#3{\hfill \begin{minipage}{1.2in}  #1
              \par\noindent #2
              \par\noindent #3
             \end{minipage}}
\renewcommand{\thefootnote}{\fnsymbol{footnote}}
%
%
\papertitlepage
\setcounter{page}{0}
\preprinumber{}{UTHEP-596}{arXiv:1001.1553}
\vspace*{2.0cm}
\begin{center}
{\large\bf\mathversion{bold}
Constant mean curvature surfaces in $AdS_3$}
\end{center}
\vskip 6ex
\baselineskip 1.0cm
\begin{center}
  {Kazuhiro ~Sakai\footnote[2]{\tt sakai@phys-h.keio.ac.jp}, }\\
\vskip -1ex
  {\it Research and Education Center for Natural Sciences}
\vskip -2ex   
  {\it and Department of Physics, Keio University}
\vskip -2ex   
  {\it Hiyoshi, Yokohama 223-8521, Japan}\\
 
\vskip 2ex
  {Yuji ~Satoh\footnote[3]{\tt ysatoh@het.ph.tsukuba.ac.jp}}\\
\vskip -1ex
  {\it Institute of Physics, University of Tsukuba}\\
\vskip -2ex
  {\it Tsukuba, Ibaraki 305-8571, Japan}

\end{center}
\vskip 13ex
%
\baselineskip=3.5ex
\begin{center} {\bf Abstract} \end{center}

We construct constant mean curvature surfaces 
of the general finite-gap type in $AdS_3$.
The special case with zero mean curvature
gives minimal surfaces relevant for
the study of Wilson loops and gluon scattering amplitudes
in ${\cal N}\!=\!4$ super Yang--Mills.
We also analyze properties of the finite-gap solutions
including asymptotic behavior 
and the degenerate (soliton) limit,
and discuss possible solutions with null boundaries.

\vskip 2ex
\vspace*{\fill}
\noindent
January 2010
\setcounter{page}{0}
\newpage
\renewcommand{\thefootnote}{\arabic{footnote}}
\setcounter{footnote}{0}
\setcounter{section}{0}
\baselineskip = 3.5ex
\pagestyle{plain}
%

\section{Introduction}

Classical string solutions in anti-de Sitter spaces
attract much attention in the study of the AdS/CFT correspondence.
Closed strings with a large spin have been of
particular interest
as they predict the strong coupling limit of
the anomalous dimension of
long operators in ${\cal N}\!=\!4$ super Yang--Mills
\cite{Gubser:2002tv}.
Recently open string solutions have also been becoming of importance
in the study of gluon scattering amplitudes
at strong coupling
\cite{Alday:2007hr,Alday:2009yn,Alday:2009dv}.
Besides the physical significance,
these classical strings possess a practical advantage
that they satisfy integrable equations of motion.
This property provides us
with a firm technical basis of quantitative investigations.

The classical integrability allows us
to construct several kinds of general class of solutions
in an explicit form.
Among others, one of the most general classes
is known as the finite-gap solutions.
Finite-gap solutions are expressed
in terms of Riemann theta functions
and contour integrals associated with an Riemann surface
called the spectral curve.
The form of general spectral curves
was studied in the case of bosonic closed strings in
$AdS_3\times S^1$\cite{Kazakov:2004nh}
and also in the case of superstrings
in $AdS_5\times S^5$ \cite{Beisert:2005bm}
and in $AdS_4\times \mathbb{CP}^3$\cite{Gromov:2008bz}.
Beyond the spectral curve, it is possible in some simple cases
to express the finite-gap solutions themselves
in a fully explicit form.
Indeed, explicit finite-gap solutions
were constructed 
in $\bbR^3, S^3, H^3$\cite{Bobenko:1991},
in $dS_3$\cite{Krichever:1994}
and also
in $S^3\times \bbR^1$\cite{Kazakov:2004qf,Dorey:2006zj,Dorey:2006mx}.

With applications to string theory in mind,
it would certainly be useful to construct general
finite-gap solutions in $AdS_3$ and clarify their general properties.
This is the aim of the present  paper. 
We focus on the case with Euclidean world-sheet,
since we are mainly motivated by the study 
of the gluon scattering amplitudes, where 
relevant solutions are mostly space-like and correspond to 
Euclidean world-sheet.
For the construction of such solutions, see
\cite{Ryang:2007bc,Jevicki:2007aa,Astefanesei:2007bk,
Dobashi:2008ia,MMT,Itoyama:2007ue,
Itoyama:2007fs,Sommerfield:2008hu,Dorn:2009kq,
Sakai:2009ut,Dorn:2009gq,
Jevicki:2009bv,Burrington:2009bh,Dorn:2009hs}.

At first sight, desired solutions would seem to
be obtained immediately by a slight modification of the above results.
Actually, there still remain several nontrivial problems.
In order to obtain fully explicit real-valued finite-gap solutions
in $AdS_3$ with a desired world-sheet signature,
one has to solve, in addition to the equations of motion,
the Virasoro constraints and the reality condition.
While the construction of solutions
to the equations of motion remains intact,
the rest conditions have to be solved case by case.
Concerning the Virasoro constraints, it is worth mentioning that 
the general finite-gap solutions in $AdS_3$ form a distinct class
among those in $AdS_3$ with extra directions, like $AdS_3\times S^1$.
The Virasoro constraints in the former case are of light-like form,
which gives rise to entirely different singularity structures of the
meromorphic differentials on the spectral curve. Therefore general
construction of finite-gap solutions in $AdS_3$ without extra
directions may as well be discussed separately.
The reality condition is intimately related to the signature
of the world-sheet.
While closed strings are considered with Minkowski world-sheet,
we need solutions with Euclidean world-sheet.
Taking these respects into account,
we need to reconsider the construction
of the finite-gap solutions
rather than to try to modify the results
obtained in the study of closed strings.

In the conformal gauge with the Virasoro constraints imposed,
the equations of motion are equivalent to the condition
that the immersion of the world-sheet describes  a minimal surface,
i.e., a surface with vanishing mean curvature.
In this paper we slightly relax the condition
and consider surfaces with a constant mean curvature.
The problem of finding constant mean curvature surfaces
is studied in detail by Bobenko \cite{Bobenko:1991}
with the target space $\bbR^3,\ S^3,\ H^3$.
Our discussion in the first half of the present paper
serves as an extension of Bobenko's construction
to the $AdS_3$ case.
In section~2 we write down the fundamental equations to solve
and see that the problem essentially reduces to
solving a variant of the elliptic Sinh-Gordon equation.
In section~3 we construct the general finite-gap solutions
to the above elliptic Sinh-Gordon equation,
with special care of the reality condition.
In section~4 we solve the rest conditions and construct 
the general finite-gap constant mean curvature surfaces in $AdS_3$.

In the latter half of the paper we investigate 
aspects of the solutions.
In section~5 we study the asymptotic behavior of
the finite-gap solutions at the boundary of the world-sheet,
which depends largely on the world-sheet signature and the reality
condition. We also consider the degenerate limit
where finite-gap solutions reduce to soliton solutions,
and discuss what kind of null boundary solutions are obtained.
This provides an extension of the search for the null boundary solutions
within the finite-gap solutions of genus one \cite{Sakai:2009ut}.
In section~6 we study the relation between our expression and
the Krichever's one constructed for $dS_{3}$ \cite{Krichever:1994}.
We conclude in section~7.

\section{Equations for constant mean curvature surfaces in $AdS_3$}

Let $\vec{Y}=(Y_{-1},Y_0,Y_1,Y_2)^{\rm T}\in\bbR^{2,2}$
denote the global coordinate parametrizing the 
$AdS_3$ spacetime. The $AdS_3$ is given as a hypersurface
\eqb\label{AdSeq}
\vec{Y}\cdot\vec{Y}:=-Y_{-1}^2-Y_0^2+Y_1^2+Y_2^2=-1
\eqe
in $\bbR^{2,2}$.
Inner product of two vectors $\vec{A},\vec{B}$
in $\bbR^{2,2}$ is defined by
\eqb
\vec{A}\cdot\vec{B}=\eta_{ab}A^aB^b
\eqe
with the metric
\eqb
\eta_{ab}=\eta^{ab}={\rm diag}(-1,-1,+1,+1).
\eqe

Let $z$ be the complex coordinate parametrizing the
Euclidean world-sheet.
We consider a map $\vec{Y}(z,\bz)$ from the complex $z$-plane
to the $AdS_3$ given by (\ref{AdSeq}).
We impose the Virasoro constraints
\eqb\label{Virasoro}
\vec{Y}_z^2=\vec{Y}_\bz^2 =0,
\eqe
where
$\vec{Y}_z = \partial_z\vec{Y},
\ \vec{Y}_\bz=\partial_\bz\vec{Y}$.
These conditions imply that the surface in consideration
is space-like.\footnote{ 
The Virasoro constraints (\ref{Virasoro}) imply that 
$\vec{Y}_s=(\vec{Y}_z+\vec{Y}_\bz)/2,\
\vec{Y}_t=i(\vec{Y}_z-\vec{Y}_\bz)/2$
are both space-like or both time-like.
As $\vec{Y}$ is time-like and
there cannot be more than two time-like directions
in $\bbR^{2,2}$, the surface must be space-like.
Then it follows that $\vec{Y}_z\cdot\vec{Y}_\bz \ge 0$.
}
For space-like surfaces one can introduce the notation
\eqb
2e^\varphi:=\vec{Y}_z\cdot\vec{Y}_\bz\,,\qquad \varphi(z,\bz)\in\bbR,
\eqe
without loss of generality.
Let us also introduce a pseudo vector
\eqb
N_a := \frac{1}{2}e^{-\varphi}\epsilon_{abcd}Y^b Y^c_z Y^d_\bz,
\qquad \epsilon_{(-1)012}=+1,
\eqe
which satisfies
\eqb
\vec{N}\cdot\vec{Y}=\vec{N}\cdot\vec{Y}_z=\vec{N}\cdot\vec{Y}_\bz=0,
\qquad
\vec{N}\cdot\vec{N}=1.
\eqe
The set of vectors $\vec{Y},\vec{Y}_z,\vec{Y}_\bz,\vec{N}$
span a basis of the moving frame in $\bbR^{2,2}$.
When writing them down into the matrix form
\eqb\label{originalphi}
\iPhi=(\vec{Y},\vec{Y}_z,\vec{Y}_\bz,\vec{N})^{\rm T},
\eqe
one can check that the following equations hold:
\eqb\label{lineqinz}
\iPhi_z=\mathcal{U}\iPhi,\qquad
\iPhi_\bz=\mathcal{V}\iPhi,
\eqe
with
\eqb\label{UVinz}
\cU=\left(\begin{array}{cccc}
0&1&0&0\\
0&\varphi_z&0&A\\
2e^\varphi&0&0&2iHe^\varphi\\
0&-iH&-\frac{1}{2}Ae^{-\varphi}&0
\end{array}\right),\quad
\cV=\left(\begin{array}{cccc}
0&0&1&0\\
2e^\varphi&0&0&2iHe^\varphi\\
0&0&\varphi_\bz&-\bA\\
0&\frac{1}{2}\bA e^{-\varphi}&-iH&0
\end{array}\right).\hspace{-2em}\nn\\
\eqe
Here $A,H$ are given by
\eqb
A:=\vec{Y}_{zz}\cdot\vec{N},\qquad
2iHe^\varphi := \vec{Y}_{z\bz}\cdot\vec{N}.
\eqe
$H$ is actually the mean curvature of the surface.
Note that
\eqb
\overline{\vec{N}}=-\vec{N},\qquad \overline{H} = H.
\eqe

As mentioned in the beginning,
classical string solutions are minimal surfaces
and thus correspond to the case
\eqb
H=0.
\eqe
Indeed, (2.9) with $H=0$
immediately leads to
the equations of motion for the classical strings
\eqb\label{stringEOM}
\vec{Y}_{z\bz}-(\vec{Y}_z\cdot\vec{Y}_\bz)\vec{Y}=0.
\eqe
In this paper, however, we consider slightly more general
situation where the mean curvature is constant,
\eqb
H={\rm const.}
\eqe
Under this restriction, the compatibility condition
\eqb
[\partial_z-\cU,\partial_\bz-\cV]=0
\eqe
yields the following equations
\eqb\label{gmshGeq}
\varphi_{z\bz}-2(1+H^2)e^\varphi+\frac{1}{2}A\bA e^{-\varphi}
\Eqn{=}0,\\
A_\bz\Eqn{=}0.
\eqe
Note that $A$ turns out to be analytic in $z$.
If we make the following change of notations,
\eqb
A(z)\Eqn{=}\delta(z)e^{\alpha(z)}=2\sqrt{1+H^2}e^{2\alpha(z)},\\
\delta(z)\Eqn{=}2\sqrt{1+H^2}e^{\alpha(z)},\\
w \Eqn{=} \int^z\delta(z) dz,\\
\hvarphi \Eqn{=} \varphi-\alpha-\balpha,
\eqe
eq.(\ref{gmshGeq}) is transformed into
a variant\footnote{
Eq. (\ref{vEllSinh}) differs from the canonical
elliptic Sinh-Gordon equation
by the sign in front of $\sinh\hvarphi$.} 
of the elliptic Sinh-Gordon equation
\eqb\label{vEllSinh}
\hvarphi_{w\bw}-\sinh\hvarphi = 0.
\eqe
We stress that $\hvarphi(w,\bw)$ has to be real.
We are going to study this equation in detail 
in the next section, where the general real finite-gap 
solutions are constructed. Those finite-gap solutions 
are single-valued on the $w$-plane, and thus
the map $w(z)$ must not have any branch points at $|z|<\infty$
in order for the surface $Y(z,\bz)$ to be single-valued on the
$z$-plane. This is in contrast with the null boundary solutions
of Alday--Maldacena \cite{Alday:2009yn}
which are single-valued on the $z$-plane but
generically have branch points in the map $w(z)$
when there are more than 4 cusps.
Therefore our finite-gap solutions
and the multi-cusp solutions generically
belong to different categories of solutions.
We will discuss some exceptional cases in section~5.

In terms of new parameters, (\ref{lineqinz}), (\ref{UVinz})
can be expressed as
\eqb\label{lineqinw}
\iPhi_w = \hU\iPhi,\qquad \iPhi_\bw = \hV\iPhi,
\eqe
with $\hU:=\delta^{-1}\cU,\ \hV:=\bdelta^{-1}\cV$,
or  explicitly
\eqb
\hU\Eqn{=}\left(\begin{array}{cccc}
0&\delta^{-1}&0&0\\
0&\varphi_w&0&e^\alpha\\
2\delta^{-1}e^\varphi&0&0&2iH\delta^{-1}e^\varphi\\
0&-iH\delta^{-1}&-\frac{1}{2}e^{-\varphi+\alpha}&0
\end{array}\right),\\[1ex]
\hV\Eqn{=}\left(\begin{array}{cccc}
0&0&\bdelta^{-1}&0\\
2\bdelta^{-1}e^\varphi&0&0&2iH\bdelta^{-1}e^\varphi\\
0&0&\varphi_\bw&-e^{\balpha}\\
0&\frac{1}{2}e^{-\varphi+\balpha}&-iH\bdelta^{-1}&0
\end{array}\right).
\eqe
The structure of the above equations becomes clearer
if one performs a suitable gauge transformation of the form
\eqb
\iPhi = g(w,\bw)\tPhi.
\eqe
In terms of $\tPhi$, equations (\ref{lineqinw}) become
\eqb\label{tlineqinw}
\tPhi_w = \tU\tPhi,\qquad \tPhi_\bw = \tV\tPhi,
\eqe
with
\eqb
\tU = g^{-1}\hU g - g^{-1}g_w,\qquad
\tV = g^{-1}\hV g - g^{-1}g_\bw.
\eqe
By taking
\eqb\label{g}
g=\left(\begin{array}{cccc}
0&ie^{ih}&e^{-ih}&0\\
0&0&0&2e^\alpha\\
2ie^{\varphi-\alpha}&0&0&0\\
0&-ie^{ih}&e^{-ih}&0
\end{array}\right)
\eqe
with
\eqb
e^{ih}=\left(\frac{1+iH}{1-iH}\right)^{\frac{1}{4}},
\eqe
one can recast $\tU,\tV$ into the tensor product form
\eqb\label{tUtV}
\tU = U(\nu)\otimes \unitmatrix + \unitmatrix \otimes U(\nu'),\qquad
\tV = V(\nu)\otimes \unitmatrix + \unitmatrix \otimes V(\nu'),
\eqe
\eqb
\nu=e^{ih},\qquad \nu'=ie^{-ih}.
\eqe
The components $U,V$ are given by
\eqb\label{UVpair}
U(\nu)=\frac{1}{2}\Biggl(\begin{array}{cc}
-\hvarphi_w & -i\nu\\
-i\nu  & \hvarphi_w
\end{array}\Biggr),\qquad
V(\nu)=\frac{i}{2\nu}\Biggl(\begin{array}{cc}
0     & e^{-\hvarphi}\\
e^\hvarphi & 0
\end{array}\Biggr),
\eqe
and $\unitmatrix$ denotes the $2\times 2$ unit matrix.
Tensor product here is understood as
\eqb
\left(\begin{array}{cc}
A_{11}&A_{12}\\
A_{21}&A_{22}
\end{array}\right)
\otimes
\left(\begin{array}{cc}
B_{11}&B_{12}\\
B_{21}&B_{22}
\end{array}\right)
=
\left(\begin{array}{cccc}
A_{11}B_{11}&A_{11}B_{12}&A_{12}B_{11}&A_{12}B_{12}\\
A_{11}B_{21}&A_{11}B_{22}&A_{12}B_{21}&A_{12}B_{22}\\
A_{21}B_{11}&A_{21}B_{12}&A_{22}B_{11}&A_{22}B_{12}\\
A_{21}B_{21}&A_{21}B_{22}&A_{22}B_{21}&A_{22}B_{22}\\
\end{array}\right).
\eqe
Due to the structure (\ref{tUtV}),
one can construct solutions to the equations (\ref{tlineqinw})
as a tensor product of more elementary ones.
Indeed, (\ref{tlineqinw}) are solved by
\eqb
\tPhi = \iPhi_0 M,\qquad
\label{Psiproduct}
\iPhi_0=\iPsi(\nu)\otimes\iPsi(\nu'),
\eqe
where $M$ is a $4\times 4$ Matrix independent of $w,\bw$
and $\iPsi(\nu)$ is a $2\times 2$ matrix
obeying the set of equations
\eqb
\label{lineqPsiinw}
\iPsi_w(\nu)=U(\nu)\iPsi(\nu),\qquad
\iPsi_\bw(\nu)=V(\nu)\iPsi(\nu).
\eqe
Note that the compatibility of the above linear equations
\eqb
[\partial_w-U,\partial_\bw-V]=0
\eqe
yields eq.~(\ref{vEllSinh}).

To summarize, once the linear problem (\ref{lineqPsiinw})
is solved, one can always construct a solution to 
the original linear equations (\ref{lineqinz}) in the form
\eqb\label{phistructure}
\iPhi=g\iPhi_0 M = g\iPsi(\nu)\otimes\iPsi(\nu') M.
\eqe
With suitably chosen $M$,
the first row of $\iPhi$ gives a real
constant mean curvature surface $Y(z,\bz)$.
We give the explicit construction of $\iPsi(\nu)$ in
the next section and then determine $M$ in Section 4.

\section{Real finite-gap solutions of
  the modified elliptic Sinh-Gordon equation}

In this section we construct
the general real finite-gap solutions to the equation
\eqb\label{vEllSinh2}
\hvarphi_{w\bw}-\sinh\hvarphi = 0,\qquad \hvarphi\in\bbR.
\eqe
This equation is a real form of the complex Sine-Gordon equation,
for which construction of the general finite-gap solutions
has been well-studied \cite{Belokolos:1994}.
The main point here is to select real valued solutions.
Such reality condition was studied in detail 
by Bobenko\cite{Bobenko:1991}
in the case of the canonical elliptic
Sinh-Gordon equation.
Since the present case differs from that case
purely by the reality condition,
one can make full use of the result in the reference.
Therefore we first summarize the result of complex valued
finite-gap solutions in \cite{Bobenko:1991}
and then study the reality condition in the present case.
We basically follow the notations in \cite{Bobenko:1991},
but some of them are modified accordingly
in order for a better fit to the present case.\footnote{
$V(\nu),R$ are modified from those in \cite{Bobenko:1991}.
}

In the standard approach to
classical integrable equations,
one starts with the auxiliary linear problem.
As mentioned in the last section,
eq.(\ref{vEllSinh2}) can be regarded as the compatibility condition
\eqb
[\partial_w-U,\partial_\bw-V]=0
\eqe
of the auxiliary linear problem
(\ref{lineqPsiinw})
with $U,V$ given in (\ref{UVpair}).
The linear differential operators
exhibit the following reduction relations
\eqb\label{LaxRedH}
\sigma_3\left(\partial_w-U(\nu)\right)\sigma_3
  \Eqn{=}\partial_w-U(-\nu),\quad\ 
\sigma_3\left(\partial_\bw-V(\nu)\right)\sigma_3
  =\partial_\bw-V(-\nu),
\eqe
\eqb\label{LaxRedA}
R^{-1}\left(\partial_w-U(\nu)\right)R
  \Eqn{=}\overline{\partial_\bw-V(\bnu^{-1})},\quad\ 
R^{-1}\left(\partial_\bw-V(\nu)\right)R
  =\overline{\partial_w-U(\bnu^{-1})},\hspace{2em}
\eqe
where
\eqb
\sigma_3 =\left(\begin{array}{cc}
1 & 0\\
0 & -1
\end{array}\right),\qquad
R=\left(\begin{array}{cc}
0       & e^{-\hvarphi/2}\\
e^{\hvarphi/2} & 0
\end{array}\right).
\eqe
These reductions induce a holomorphic involution
$\nu\to -\nu$ and an antiholomorphic involution
$\nu\to\bnu^{-1}$ of the solutions.
In particular, the latter determines the reality condition
of the present system, which we will discuss later.

Finite-gap solutions are characterized by the spectral curve.
In the present case, it is a Riemann surface of
the hyperelliptic curve $C$ defined by
\eqb
C:\quad \tmu^2=\lambda\prod_{l=1}^{2g}(\lambda-\lambda_l).
\eqe
Note that it is parametrized by $\lambda:=\nu^2$,
due to the invariance under the involution $\nu\to -\nu$.
Let
$a_n,\,b_n,\,n=1,\ldots,g$
denote the canonical basis of the 1-cycles
and $du_n$ the basis of holomorphic abelian differentials.
One can take $du_n$ satisfying the normalization conditions
\eqb
\oint_{a_m}du_n = 2\pi i\delta_{mn}.
\eqe
The associated period matrix is given by
\eqb\label{periodMat}
B_{mn}=\oint_{b_m}du_n.
\eqe
This defines the Riemann theta function
\eqb\label{RiemannTheta}
\theta(\vz)=\sum_{\vm\in\bbZ^g_{\phantom{g}}}\exp\left(\frac{1}{2}
\langle \vm,B\vm\rangle + \langle \vz,\vm\rangle\right),
\qquad \vz \in\bbC^g,
\eqe
where $\langle \cdot, \cdot \rangle$ denotes
$g$-dimensional Euclidean inner-product.

In order to express
the solutions of (\ref{lineqPsiinw}),
let us introduce the Riemann surface $\hC$
parametrized by $\nu$. $\hC$ is a double cover of $C$,
consisting of two branches $\nu=\pm\sqrt{\lambda}$.
We take a closed path
\eqb
{\cal L}\in[a_1+\cdots+a_g]
\eqe
as the branch cut connecting the two branches.
With slight abuse of notation we will use the same symbols
for contours on $C$ and their lift on $\hC$.
We also introduce
abelian differentials of the second kind
$d\iOmega_i\ (i=1,2)$
fixed by the normalization condition
\eqb
\oint_{a_n}d\iOmega_i=0,\qquad i=1,2,\qquad n=1,\ldots,g
\eqe
and the asymptotic behavior
\eqb
d\iOmega_1\Eqn{\to} d\nu,\phantom{-,}    \qquad  \nu\to\infty,\\
d\iOmega_2\Eqn{\to} -\frac{d\nu}{\nu^2},\qquad  \nu\to 0.
\eqe

Now, the general complex finite-gap solutions of eq.(\ref{vEllSinh2})
can be expressed as
\eqb
\hvarphi = 2\log\frac{\theta(\vX)}{\theta(\vX+\vDelta)}
\eqe
where
\[
\vX = -\frac{i}{2}(\vU w - \vV \bw) + \vD,
\]
\[
U_n = \oint_{b_n}d\iOmega_1,\qquad
V_n = \oint_{b_n}d\iOmega_2,
\]
\[
\vDelta=\pi i(1,\ldots,1),
\]
and $\vD\in \bbC^g$ is a constant.
The corresponding solution to eq.(\ref{lineqPsiinw}),
namely the Baker--Akhiezer function, is given by
\eqb
\iPsi(\nu)
\Eqn{=}\left(\begin{array}{cc}
\psi_1(\nu) & \psi_1^\ast(\nu)\\
\psi_2(\nu) & \psi_2^\ast(\nu)
\end{array}\right)
=\left(\begin{array}{cc}
 \frac{\theta(\vu+\vX)\theta(\vD)}
      {\theta(\vu+\vD)\theta(\vX)}e^\omega &
 \frac{\theta(-\vu+\vX)\theta(\vD)}
      {\theta(-\vu+\vD)\theta(\vX)}e^{-\omega} \\
 \frac{\theta(\vu+\vX+\vDelta)\theta(\vD)}
      {\theta(\vu+\vD)\theta(\vX+\vDelta)}e^\omega &
-\frac{\theta(-\vu+\vX+\vDelta)\theta(\vD)}
      {\theta(-\vu+\vD)\theta(\vX+\vDelta)}e^{-\omega}
\end{array}\right),\quad
\eqe
where
\[
\omega=-\frac{i}{2}\left(\iOmega_1 w - \iOmega_2 \bw\right),
\]
\eqb\label{uofnu}
\vu(\nu) = \int_\ell d\vu,\quad
\iOmega_i(\nu) = \int_\ell d\iOmega_i.
\eqe
Here $\ell$ is a path which connects
points $\infty$ and $\nu$ without intersecting $\cL$.
The above Baker--Akhiezer function
transforms under the holomorphic involution as
\eqb\label{BARedH}
\iPsi(-\nu)=\sigma_3\iPsi(\nu)\sigma_1,
\eqe
which is compatible with the reduction (\ref{LaxRedH}).
The construction so far is identical to that
in \cite{Bobenko:1991},
as we have regarded $w$ and $\bw$ as independent variables.\footnote{
As far as complex conjugation does not concern,
eq.(\ref{vEllSinh2}) is identified with
the canonical elliptic Sinh-Gordon equation
by the transformation $(w,\bw)\mapsto(w,-\bw)$.}

We are now in a position to consider the reality condition.
{}From now on we set $w$ and $\bw$ complex conjugate
to each other.
The reality of $\hvarphi$, combined with the reduction (\ref{LaxRedA}),
induces an antiholomorphic involution $\nu\to \bnu^{-1}$.
It appears on the spectral curve $C$ as
\eqb
\tau:\quad \lambda\to \overline{\lambda}^{-1}.
\eqe
In order for $C$ to have this property,
we consider $C$ with all the branch points $\lambda_i$
divided into pairs
\eqb
\overline{\lambda}_{2n-1}=\lambda^{-1}_{2n},\quad n=1,\ldots,g,\qquad
|\lambda_i|\ne 1.
\eqe
The canonical basis $a_n,\,b_n$ can be chosen
so that they transform under the involution as
\eqb
\tau a_n = -a_n,\qquad
\tau b_n = b_n - a_n + \sum_{i=1}^g a_i\,.
\eqe
With this basis, the period matrix has the form 
\eqb\label{Bmn}
B_{mn}=B^{\rm R}_{mn}+\pi i(1-\delta_{mn}),
\qquad B^{\rm R}_{mn}<0
\eqe
and the theta function exhibits
the following simple conjugation property
\eqb
\overline{\theta(\vz)}=\theta(\overline{\vz}).
\eqe
If we fix $\cal{L}$ so that
\eqb
\tau{\cal L}={\cal L},
\eqe
the anti-holomorphic involution $\tau^\ast$ on $\hC$ is
realized as
\eqb
\overline{\tau^\ast \nu}=\nu^{-1}.
\eqe
With these choice of cycles and branches,
we see that
\eqb
\tau^\ast d\iOmega_1=\overline{d\iOmega_2}
\eqe
and thus
\eqb
\overline{\vU}=\vV.
\eqe
One can also show that
\eqb\label{Imvu}
  \overline{\vu(\nu)} = \vu(\overline{\nu}^{-1}) - \vDelta  
  \quad \operatorname{mod}\ 2\pi i\bbZ^g .
\eqe

Now observe that
\eqb
\overline{\hvarphi}
\Eqn{=}\overline{2\log\frac{\theta(\vX)}{\theta(\vX+\vDelta)}}
=2\log\frac{\theta(\overline{\vX})}
           {\theta(\overline{\vX}-\vDelta)}
=2\log\frac{\theta(\overline{\vX})}
           {\theta(\overline{\vX}+\vDelta)}.
\eqe
Taking account of the periodicity
$\theta(\vz+2\pi i\vm)=\theta(\vz),\ \vm\in\bbZ^g$,
one finds that $\hvarphi$ is real if $\vX$ is real mod $\pi i$.
Since $\overline{\vU}=\vV$, this condition is equivalent
to the condition that $\vD$ is real mod $\pi i$
\eqb
\vD\in\bbR^g\ \operatorname{mod}\ \pi i\bbZ^g.
\eqe

Next let us see how the antiholomorphic involution
acts on the Baker--Akhiezer function.
Let us introduce the following quantities
\[
a_\pm(\nu) :=
\frac{\theta(\vu(\nu)\pm \vD)}{\theta(\vu(\nu)\pm \vD+\vDelta)},
\]
\[
\label{dexpression}
d:=\psi_1\psi_2^\ast-\psi_1^\ast\psi_2
=-2\frac{\theta^2(\vD)}{\theta(\vzero)\theta(\vDelta)}\cdot
\frac{\theta(\vu)\theta(\vu+\vDelta)}{\theta(\vu-\vD)\theta(\vu+\vD)}.
\]
One can verify that
\[
\label{alphaconjugate}
\overline{a_\pm(\nu)}=a_\pm^{-1}(\bnu^{-1}),
\]
\[
\label{dconjugate}
\overline{d(\nu)}
=d(\bnu^{-1})a_+(\bnu^{-1})a_-(\bnu^{-1}).
\]
The Baker--Akhiezer function satisfies the following conjugation
relation
\eqb\label{BARedA}
\overline{\iPsi(\nu)}
\Eqn{=}
R\iPsi(\bnu^{-1})
\left(\begin{array}{cc}
a_+(\bnu^{-1}) & 0\\
0                   & -a_-(\bnu^{-1}) 
\end{array}\right).
\eqe
This is compatible with the reduction relation (\ref{LaxRedA}).

\section{The general finite-gap solutions in $AdS_3$}

We now have the general finite-gap solutions to
the set of linear equations (\ref{lineqinz}).
In order to have real constant mean curvature surfaces,
$\iPhi$ has to satisfy
in addition the correct normalization condition
and the reality condition.
Knowing the explicit form of $\iPsi(\nu)$,
one can evaluate these conditions
and determine the form of $M$ in (\ref{phistructure}).

We use abbreviated notation
$
a_\pm=a_\pm(\nu),\,a_\pm'=a_\pm(\nu'),\,
d=d(\nu),\,d'=d(\nu')$ below.
It it is helpful to note that
$|\nu|=|\nu'|=|a_\pm|=|a_\pm'|=1$.

By definition, $\iPhi$ has to satisfy the following
normalization condition
\eqb
\iPhi
\left(\begin{array}{cccc}
-1&&&\\
&-1&&\\
&&1&\\
&&&1
\end{array}\right)
\iPhi^{\rm T}
=
\left(\begin{array}{cccc}
-1&&&\\
&0&2e^\varphi&\\
&2e^\varphi&0&\\
&&&1
\end{array}\right).
\eqe
It can be translated into the condition for $M$ as
\eqb\label{Mnorm}
M
\left(\begin{array}{cccc}
-1&&&\\
&-1&&\\
&&1&\\
&&&1
\end{array}\right)
M^{\rm T}
\Eqn{=}\frac{i}{2dd'}
\left(\begin{array}{cccc}
&&&-1\\
&&1&\\
&1&&\\
-1&&&
\end{array}\right).
\eqe
Next let us see the reality condition
\eqb
\overline{\iPhi}
=
\left(\begin{array}{cccc}
1&&&\\
&0&1&\\
&1&0&\\
&&&-1
\end{array}\right)
\iPhi.
\eqe
With the knowledge of conjugation properties
studied in the last section,
it can be written in terms of $M$ as
\eqb\label{Mreality}
\overline{M}
\Eqn{=}i\left(\begin{array}{cccc}
\left(a_+a_+'\right)^{-1}&&&\\
&-\left(a_+a_-'\right)^{-1}&&\\
&&-\left(a_-a_+'\right)^{-1}&\\
&&&\left(a_-a_-'\right)^{-1}
\end{array}\right)M.
\eqe

The matrix satisfying the equations (\ref{Mnorm}), (\ref{Mreality})
is found to be
\eqb\label{Msolution}
M\Eqn{=}\frac{1}{2|b|^{\frac{1}{2}}}
\left(\begin{array}{cccc}
e^{-\frac{\pi i}{4}}
 \left(a_+a_+'\right)^{\frac{1}{2}}&&&\\
&e^{\frac{\pi i}{4}}
 \left(a_+a_-'\right)^{\frac{1}{2}}&&\\
&&e^{\frac{\pi i}{4}}
 \left(a_-a_+'\right)^{\frac{1}{2}}&\\
&&&e^{-\frac{\pi i}{4}}
 \left(a_-a_-'\right)^{\frac{1}{2}}
\end{array}\right)\nn\\
&&\times\left(\begin{array}{cccc}
1&&&1\\
&1&1&\\
&-\sigma&\sigma&\\
-\sigma&&&\sigma
\end{array}\right)
\iLambda,
\eqe
where 
\eqb
b(\nu,\nu')
= dd'\left(a_+a_-a_+'a_-'\right)^{\frac{1}{2}}
\in\bbR,
\qquad
\sigma={\rm sign}(b),
\eqe
and 
$\iLambda\in\grp{O}(2,2)$ is an arbitrary constant matrix.
With this $M$ and $\iPsi(\nu)$ obtained in section 3,
(\ref{phistructure}) gives the general finite-gap solutions
for the real constant mean curvature surfaces in $AdS_3$.

Finally let us see the case $h=0$, where the surfaces
become minimal and give the string solutions.
For simplicity we set $\iLambda=1$.
Collecting the results so far, the immersion
of the surfaces is given by
\eqb\label{immersion}
\begin{array}{rclrcl}
Y_{-1}+Y_{2} \Eqn{=}  
  c_{\vu,\vu'}\, I_{\vu,\vu'}\, e^{\omega+\omega'}, &
Y_{-1}-Y_{2} \Eqn{=} 
  \sigma c_{-\vu,-\vu'}\,I_{-\vu,-\vu'}\,e^{-(\omega+\omega')},\\[1ex]
Y_{0}+Y_{1} \Eqn{=}
  c_{-\vu',\vu}\, I_{-\vu',\vu}\, e^{\omega-\omega'}, &
Y_{0}-Y_{1}  \Eqn{=}
  \sigma c_{\vu',-\vu}\, I_{\vu',-\vu}\, e^{-(\omega-\omega')} ,  
\end{array}
\eqe
where
\eqb
 c_{\vu,\vu'} \Eqn{=}
 \left( \frac{1}{|b|}
\frac{\theta^4(\vD)}
  {\theta(\vu+\vD)\theta(\vu+\vD+\vDelta)
   \theta(\vu'+\vD)\theta(\vu'+\vD+\vDelta)}\right)^{\frac{1}{2}},\\[1ex]
 I_{\vu,\vu'} \Eqn{=} 
\frac{e^{\frac{\pi i}{4}}\theta(\vu + \vX) \theta(\vu'+\vX+\vDelta) 
     +e^{-\frac{\pi i}{4}}\theta(\vu+\vX+\vDelta) \theta(\vu' + \vX)}
     {\theta(\vX)\theta(\vX+\vDelta)}.
\eqe
Here $\vu,\vu'$ as well as $\omega,\omega'$ are
the quantities evaluated at $\nu=1,\nu=i$, respectively.

\section{General properties and the degenerate limit}
In this section we discuss properties of the solutions. 
First, we note that
the coordinates $Y_a$ given in the form (\ref{immersion})
are oscillating with sign changes:
Since
$\theta(\vz + B \vn)
= \exp\bigl[ - \frac{1}{2} \langle \vn,
B\vn \rangle - \langle \vz, \vn \rangle 
\bigr] \theta(\vz)$ with $\vn \in \bbZ^{g}$,
$I_{\vu,\vu'}$ obtains a factor $e^{-\langle \vn, \vu+\vu'\rangle}$
under $\vX \to \vX + B\vn$.
Note that the imaginary part of $\vu+\vu'$ is
$\vDelta$ (mod $2 \pi i \bbZ^{g}$). 
Then, when the spectral curve has an odd genus and
each component of $\vn$ is, e.g., $\pm 1$, 
$B\vn$ is real (mod $2 \pi i \bbZ^{g}$) and $I_{\vu, \vu'}$
changes the sign.
This implies that 
$Y_{a}$ change the sign as $w$ and hence $\vX$ vary. 
By analyzing the zeros of the theta function,
one can also observe the sign change
in the case of even genus.
In the above discussion,
it is important that ${\rm Im}(\vu+\vu') = \vDelta/i$, $\vX$ is real
(mod $\pi i \bbZ^g$) and $B$ has a particular form given
in (\ref{Bmn}), 
which is essential for the reality of the surfaces.

For the application to the gluon scattering amplitudes, string solutions
with null boundaries without such oscillations are needed. 
Since the oscillation may disappear in the limit where the period matrix
$B$ diverges, 
we discuss such a limit in the following. 

A way to realize such a limit is to make the spectral curve degenerate.
As the simplest one, we consider the limit achieved by
shrinking every branch cut to a point on the unit circle:
\eqb\label{deglim}
 \lambda_{2n-1}, \lambda_{2n} (=\bar{\lambda}^{-1}_{2n-1} )
  \to e^{2i\phi_{n}}:= \nu_{n}^{2} \quad (n= 1, ..., g) .
\eqe
In general, this type of degeneration gives
soliton solutions\cite{Belokolos:1994}.\footnote{
For applications of soliton solutions to AdS strings,
see \cite{Jevicki:2009uz} and references therein.
}
In this limit, the geometric quantities describing the solutions are
simplified.
The abelian differentials become 
\[
 du_{n} = \frac{\nu_{n}d\lambda}{\sqrt{\lambda}(\lambda-\nu_{n}^{2})},
\]
\[ 
 d\iOmega_{1} = d\nu, \quad d\iOmega_{2}  = d\nu^{-1} ,
\]
and hence
\eqb\label{un}
u_{n} = \log \frac{\nu - \nu_{n}}{\nu + \nu_{n}}, 
\eqe
\[
 \iOmega_{1} = \nu , \quad \iOmega_{2} = \frac{1}{\nu}, 
\]
\[
 U_{n} = \overline{V}_n = 2 \nu_n.
\]
The period matrix is also given by 
\eqb
 B_{mn} = \log \Bigl(\frac{\nu_{m}-\nu_{n}}{\nu_{m}+\nu_{n}}\Bigr)^{2}.
\eqe
The diagonal elements become divergent: $B_{nn}\to-\infty$.
As for $\vD$, we set
\eqb
  D_{n} = - \frac{1}{2} B_{nn} + \eta_{n} ,
\eqe
with real (mod $\pi i $) and finite $\eta_n$. 
The theta function then reduces to a finite sum:
\eqb
\theta(\vz+\vD) \to \sum_{m_{k} =0,1}
\exp\Bigl[ \sum_{k>l} m_{k} B_{kl} m_{l} 
   + \sum_{k} m_{k}(z+\eta)_{k}\Bigr] .
\eqe

Now we are ready to discuss the string solutions in the degenerate
limit. First, let us consider the case where the original spectral curve
is of genus one. To analyze the solutions, it is useful to note that,
for example, 
$I_{u,u'}$  factorizes as 
$I_{u,u'} \sim (1-e^{u+u'+\rho})/(1-e^{\rho})$
for $u, u'$ given in (\ref{un})
and thus
\eqb\label{YFact}
 Y_{-1}+Y_{2} \sim  
  \frac{\theta(\vu+\vu'+\vX + \vDelta)}{\theta(\vX+\vDelta)}
  e^{\frac{1}{2}(s+t)}
\eqe
up to a numerical factor.
Here,
\[
  \rho_n = X_n + \frac{1}{2} B_{nn} ,
\]
and we have omitted the subscript $n$. 
One then finds that the solutions reduce to those obtained in
\cite{Sakai:2009ut}, so that
$Y_{-1} + Y_2 \sim [\cos \theta -
\frac{1}{\sqrt{2}} \tanh \frac{\rho+\pi i}{2}]
e^{\frac{1}{2}(s+t)} $ 
with $w = \frac{1}{2}(s + it)$
and $\theta$ a constant.
In the solutions, the unwanted oscillation is in fact absent.
When ${\rm Im}(\rho+\pi i) =0$,
the solutions describe the surfaces with six null boundaries extending 
to the boundary of $AdS_{3}$.
When Im$\, \rho =0$,\footnote{
This case is obtained by shifting $B$ 
in \cite{Sakai:2009ut} by $\pi i/2$.} 
the zeros of $\cosh \frac{\rho+\pi i}{2}$ are mapped to the AdS
boundary. Thus the surfaces have non-null boundaries in addition to 
the null boundaries which are the image of the world-sheet boundary
$|w| \to \infty$.

Let us move on to the case where the spectral curve has $g = 2$.
Similarly to the $g=1$ case,
$I_{\vu,\vu'}$ factorizes and the solutions
take the form (\ref{YFact}).
The surfaces then have null boundaries coming from the world-sheet
boundary. However, since
\eqb
   \theta(\vX+\vDelta) = 1 - e^{\rho_{1}} - e^{\rho_{2}}+ e^{B_{12} 
    + \rho_1+\rho_2} ,
\eqe
and $e^{B_{12}} < 0$, the denominator of $Y_{-1}+Y_{2}$ vanishes
at internal points of the world-sheet.
These zeros give non-null boundaries,
as in the case of $g=1$ with Im$\, \rho =0$.

For $g\geq 3$, because of the factors $e^{B_{mn}} < 0$ $(m \neq n)$,
the surfaces similarly have both null and non-null boundaries.
The number of the null boundaries can be increased according to $g$.
We find that the case of $g=1$ is special in that the degenerate limit
(\ref{deglim}) gives the string solutions with only null boundaries.

\section{Expression respecting the vector representation}

In general, an auxiliary linear problem to
integrable string equations of motion
admits several representations.
In the case of a constant Ricci curvature target space, it
can be expressed either in the spinor representation or in the vector
representation of the orthogonal isometry group\cite{Beisert:2004ag}.
The construction in this paper is based on the spinor representation.
This results in the tensor product structure of the solutions.
On the other hand, it should certainly be possible to
express the solution in a form respecting the vector representation.
Indeed, Krichever's construction
for the $dS_3$ case \cite{Krichever:1994} was of this form.
In this section we will see how our solutions
are related to the Krichever's.
For simplicity we restrict ourselves to the genus-one case,
but the discussion can be generalized to the
case of arbitrary genus.

At $g=1$, the Riemann theta function (\ref{RiemannTheta})
reduces to the elliptic theta function
\eqb
\theta(z)=\theta_3(z|B),
\eqe
with $B<0$.
In this section we adopt a slightly modified convention
for the Jacobi theta functions, denoted by
$\theta_j(z|B),\ j=1,2,3,4$.
The canonical Jacobi theta functions $\varth_j$
are expressed in terms of $\theta_j$ as
\eqb
\varth_j(z|\tau)=\theta_j(2\pi iz|2\pi i\tau),\qquad j=1,2,3,4.
\eqe
Below we abbreviate $\theta_j(z|B)$ as $\theta_j(z)$.

The spectral curve $\hC$
compatible with the reality condition
takes the form
\eqb
\tmu^2=\nu^2(\nu^2-\nu_1^2)(\nu^2-\overline{\nu}_1^{-2}).
\eqe
This curve can be uniformized as
\eqb
\label{nuofu}
\nu\Eqn{=}ie^{i\gamma}\frac{\theta_2(u)}{\theta_1(u)},
\qquad
\nu_1=e^{i\gamma}\frac{\theta_4(0)}{\theta_3(0)},\\[1ex]
\tmu\Eqn{=}ie^{3i\gamma}
\frac{\theta_2(0)^2\theta_2(u)\theta_3(u)\theta_4(u)}
     {\theta_3(0)\theta_4(0)\theta_1(u)^3}.
\eqe
Here $\gamma\in \bbR$ denotes the phase of $\nu_1$.
(\ref{nuofu}) inversely expresses the relation (\ref{uofnu}).
Below we let $u,u'$ denote $u(\nu)$
evaluated at $\nu=1,\nu=i$, respectively.
It then follows from (\ref{nuofu}) that
\eqb\label{identitynu1i}
i\frac{\theta_2(u)}{\theta_1(u)}=\frac{\theta_2(u')}{\theta_1(u')}.
\eqe

Now one can show that our solutions can be transformed
into the Krichever's form.
The essential difference of the two expressions is
the world-sheet coordinate dependence through theta functions,
which is collected in $I_{\vu,\vu'}$.
At $g=1$ it can be expressed as
\eqb
I_{u,u'}\Eqn{=}
\frac{ e^{\frac{\pi i}{4}}\theta_3(u+X)\theta_4(u'+X)
      +e^{-\frac{\pi i}{4}}\theta_4(u+X)\theta_3(u'+X)}
     {\theta_3(X)\theta_4(X)}.
\eqe
By using the following theta function identity
\eqb
&&\hspace{-3em}
\theta_1(z-w)\theta_2(0)\theta_3(X)\theta_4(z+w+X)\nn\\
\Eqn{=}\theta_1(z)\theta_2(w)\theta_4(z+X)\theta_3(w+X)
      -\theta_2(z)\theta_1(w)\theta_3(z+X)\theta_4(w+X)
\eqe
and (\ref{identitynu1i}),
one obtains
\eqb
I_{u,u'}=e^{-\frac{\pi i}{4}}
\frac{\theta_1(u-u')\theta_2(0)}
{\theta_1(u)\theta_2(u')}
\frac{\theta_4(u+u'+X)}{\theta_4(X)}.
\eqe
One finds that
$w$-dependence now appears only through $\theta_4$'s.
This is the form obtained in \cite{Sakai:2009ut},
and gives an $AdS_{3}$ analog 
of the form found in \cite{Krichever:1994} for $dS_{3}$.
We note that our construction however uses only
the ordinary Riemann theta function instead of 
the Prym theta function.
The above result also shows that 
the factorized form (\ref{YFact}) in the degenerate limit 
generally holds for $g=1$. The factorization for $g=2$ found there
may be regarded as a consequence of a generalization of the discussion
here to the case of arbitrary genus.

\section{Conclusions}

We have constructed space-like constant mean curvature
surfaces of the general finite-gap type in $AdS_{3}$.
In a special case with vanishing mean curvature,
our results provide a fully explicit form of
the general finite-gap open string solutions 
in $AdS_{3}$ with Euclidean world-sheet.
The points in the construction are
concerning the reality condition and the Virasoro constraints
without extra directions.
These are not achieved by a simple modification of the results
\cite{Krichever:1994,Kazakov:2004qf,Dorey:2006zj,Dorey:2006mx},
but rather need analysis analogous to that in \cite{Bobenko:1991}.

We have also analyzed properties of the solutions. 
Generically, the solutions oscillate with sign changes.
In a degenerate (soliton) limit, the solutions describe
the surfaces with null boundaries 
coming from the world-sheet boundary.
For the genus of the spectral curve $g \geq 2$,
the surfaces in addition have other type of boundaries mapped from
internal part of the world-sheet. The case of $g=1$ is special
in that it has solutions with only null boundaries.

The analysis here is an extension of the search of the null boundary
solutions among finite-gap solutions carried out for $g=1$
in \cite{Sakai:2009ut}.
The solutions in this paper generically have different analytic
structures from those for the null boundary solutions
of Alday--Maldacena\cite{Alday:2009yn},
and  thus belong to a different class.
We have also discussed the relation to the construction 
for $dS_{3}$ \cite{Krichever:1994} based on the Prym theta function.

The construction in this paper can be extended to
the string solutions in $AdS_{3}$ with Minkowski world-sheet,
by reanalyzing the antiholomorphic reduction condition
(\ref{LaxRedA}) and the resulting reality conditions accordingly.
The large class of
classical solutions from both Euclidean and Minkowski world-sheets 
may shed light on the study of strings in $AdS_{3}$.
In particular, finding non-oscillating solutions with appropriate
boundaries would be useful for exploring the scattering amplitudes and
the Wilson loops in gauge theories.

\vspace{3ex}

\begin{center}
  {\bf Acknowledgments}
\end{center}

We would like to thank A.~Kato and K.~Mohri for useful discussions.
The work of K.S. and Y.S. is supported in part by Grant-in-Aid for
Scientific Research from the Japan Ministry of Education, Culture, 
Sports, Science and Technology.

%
%
\def\thebibliography#1{\list
 {[\arabic{enumi}]}{\settowidth\labelwidth{[#1]}\leftmargin\labelwidth
  \advance\leftmargin\labelsep
  \usecounter{enumi}}
  \def\newblock{\hskip .11em plus .33em minus .07em}
  \sloppy\clubpenalty4000\widowpenalty4000
  \sfcode`\.=1000\relax}
 \let\endthebibliography=\endlist
\vspace{3ex}
\begin{center}
 {\bf References}
\end{center}
\par \vspace*{-2ex}


\begin{thebibliography}{999}
\parskip=-2.5pt
%

\bibitem{Gubser:2002tv}
  S.~S.~Gubser, I.~R.~Klebanov and A.~M.~Polyakov,
  Nucl.\ Phys.\  B {\bf 636} (2002) 99
  [arXiv:hep-th/0204051].

\bibitem{Alday:2007hr}
  L.~F.~Alday and J.~M.~Maldacena,
  JHEP {\bf 0706} (2007) 064
  [arXiv:0705.0303 [hep-th]].

\bibitem{Alday:2009yn}
  L.~F.~Alday and J.~Maldacena,
  JHEP {\bf 0911} (2009) 082
  [arXiv:0904.0663 [hep-th]].

\bibitem{Alday:2009dv}
  L.~F.~Alday, D.~Gaiotto and J.~Maldacena,
  arXiv:0911.4708 [hep-th].

\bibitem{Kazakov:2004nh}
  V.~A.~Kazakov and K.~Zarembo,
  JHEP {\bf 0410} (2004) 060
  [arXiv:hep-th/0410105].

\bibitem{Beisert:2005bm}
  N.~Beisert, V.~A.~Kazakov, K.~Sakai and K.~Zarembo,
  Commun.\ Math.\ Phys.\  {\bf 263} (2006) 659
  [arXiv:hep-th/0502226].

\bibitem{Gromov:2008bz}
  N.~Gromov and P.~Vieira,
  JHEP {\bf 0902} (2009) 040
  [arXiv:0807.0437 [hep-th]].

\bibitem{Bobenko:1991}
  A.~I.~Bobenko,
  Math.\ Ann. {\bf 290} (1991) 209.

\bibitem{Krichever:1994}
  I.~M.~Krichever,
  Funct.\ Anal.\ Appl.\ {\bf 28} (1994) 21.

\bibitem{Kazakov:2004qf}
  V.~A.~Kazakov, A.~Marshakov, J.~A.~Minahan and K.~Zarembo,
  JHEP {\bf 0405} (2004) 024
  [arXiv:hep-th/0402207].

\bibitem{Dorey:2006zj}
  N.~Dorey and B.~Vicedo,
  JHEP {\bf 0607} (2006) 014
  [arXiv:hep-th/0601194].

\bibitem{Dorey:2006mx}
  N.~Dorey and B.~Vicedo,
  JHEP {\bf 0703} (2007) 045
  [arXiv:hep-th/0606287].

\bibitem{Ryang:2007bc}
  S.~Ryang,
  Phys.\ Lett.\  B {\bf 659} (2008) 894
  [arXiv:0710.1673 [hep-th]].

\bibitem{Jevicki:2007aa}
  A.~Jevicki, K.~Jin, C.~Kalousios and A.~Volovich,
  JHEP {\bf 0803} (2008) 032
  [arXiv:0712.1193 [hep-th]].

\bibitem{Astefanesei:2007bk}
  D.~Astefanesei, S.~Dobashi, K.~Ito and H.~Nastase,
  JHEP {\bf 0712} (2007) 077
  [arXiv:0710.1684 [hep-th]].

\bibitem{Dobashi:2008ia}
  S.~Dobashi, K.~Ito and K.~Iwasaki,
  JHEP {\bf 0807} (2008) 088
  [arXiv:0805.3594 [hep-th]].

\bibitem{MMT}
  A.~Mironov, A.~Morozov and T.~N.~Tomaras,
  JHEP {\bf 0711} (2007) 021
  [arXiv:0708.1625 [hep-th]]; Phys.\ Lett.\  B {\bf 659} (2008) 723
  [arXiv:0711.0192 [hep-th]].

\bibitem{Itoyama:2007ue}
  H.~Itoyama, A.~Mironov and A.~Morozov,
  Nucl.\ Phys.\  B {\bf 808} (2009) 365
  [arXiv:0712.0159 [hep-th]].

\bibitem{Itoyama:2007fs}
  H.~Itoyama and A.~Morozov,
  Prog.\ Theor.\ Phys.\  {\bf 120} (2008) 231
  [arXiv:0712.2316 [hep-th]].

\bibitem{Sommerfield:2008hu}
  C.~M.~Sommerfield and C.~B.~Thorn,
  Phys.\ Rev.\  D {\bf 78} (2008) 046005
  [arXiv:0805.0388 [hep-th]].

\bibitem{Dorn:2009kq}
  H.~Dorn, G.~Jorjadze and S.~Wuttke,
  JHEP {\bf 0905} (2009) 048
  [arXiv:0903.0977 [hep-th]].

\bibitem{Sakai:2009ut}
  K.~Sakai and Y.~Satoh,
  JHEP {\bf 0910} (2009) 001
  [arXiv:0907.5259 [hep-th]].

\bibitem{Dorn:2009gq}
  H.~Dorn,
  JHEP {\bf 1002} (2010) 013
  [arXiv:0910.0934 [hep-th]].
  
\bibitem{Jevicki:2009bv}
  A.~Jevicki and K.~Jin,
  arXiv:0911.1107 [hep-th].
 
\bibitem{Burrington:2009bh}
  B.~A.~Burrington and P.~Gao,
  arXiv:0911.4551 [hep-th].
  
\bibitem{Dorn:2009hs}
  H.~Dorn, N.~Drukker, G.~Jorjadze and C.~Kalousios,
  arXiv:0912.3829 [hep-th].

\bibitem{Belokolos:1994}
  E.~D.~Belokolos, A.~I.~Bobenko, V.~Z.~Enolski, A.~R.~Its
  and V.~B.~Matveev,
  ``Algebro-geometric approach in the theory of integrable equations'',
Springer Series in Nonlinear Dynamics, Springer, Berlin (1994).

\bibitem{Jevicki:2009uz}
  A.~Jevicki and K.~Jin,
  JHEP {\bf 0906} (2009) 064
  [arXiv:0903.3389 [hep-th]].
  
\bibitem{Beisert:2004ag}
  N.~Beisert, V.~A.~Kazakov and K.~Sakai,
  Commun.\ Math.\ Phys.\  {\bf 263} (2006) 611
  [arXiv:hep-th/0410253].

%
\end{thebibliography}
\end{document}